\documentclass[prl,aps,10pt,superscriptaddress,twocolumn,floatfix]{revtex4-1}
\usepackage{graphicx,color}
\usepackage[nice]{nicefrac}							
\usepackage{amsmath,amssymb,bm}

\usepackage[plainpages=false,pdfpagelabels,colorlinks=true,linkcolor=red,urlcolor=blue,citecolor=blue,pdftitle={},pdfauthor={},pdfdisplaydoctitle=true,pdfduplex=DuplexFlipLongEdge]{hyperref}

\definecolor{darkred}{rgb}{0.90,0.2,0.2}
\definecolor{darkgreen}{rgb}{0,0.60,.2}
\definecolor{darkblue}{rgb}{0.1,0.3,1}
\definecolor{grey}{cmyk}{0,0,0,0.25}
\definecolor{orange}{cmyk}{0,0.6,0.8,0}

\begin{document}

\title{Volume Law and Quantum Criticality in the \\ Entanglement Entropy of Excited Eigenstates of the Quantum Ising Model}

\author{Lev Vidmar}
\affiliation{Department of Theoretical Physics, J. Stefan Institute, SI-1000 Ljubljana, Slovenia}
\affiliation{Kavli Institute for Theoretical Physics, University of California, Santa Barbara, California 93106, USA}
\author{Lucas Hackl}
\affiliation{Max  Planck  Institute of Quantum Optics, Hans-Kopfermann-Stra\ss e 1, D-85748 Garching bei M\"unchen, Germany}
\affiliation{Department of Physics, The Pennsylvania State University, University Park, PA 16802, USA}
\affiliation{Institute for Gravitation and the Cosmos, The Pennsylvania State University, University Park, PA 16802, USA}
\author{Eugenio Bianchi}
\affiliation{Department of Physics, The Pennsylvania State University, University Park, PA 16802, USA}
\affiliation{Institute for Gravitation and the Cosmos, The Pennsylvania State University, University Park, PA 16802, USA}
\author{Marcos Rigol}
\affiliation{Department of Physics, The Pennsylvania State University, University Park, PA 16802, USA}
\affiliation{Kavli Institute for Theoretical Physics, University of California, Santa Barbara, California 93106, USA}


\begin{abstract}
Much has been learned about universal properties of entanglement entropies in ground states of quantum many-body lattice systems. Here we unveil universal properties of the average bipartite entanglement entropy of eigenstates of the paradigmatic quantum Ising model in one dimension. The leading term exhibits a volume-law scaling that we argue is universal for translationally invariant quadratic models. The subleading term is constant at the critical field for the quantum phase transition and vanishes otherwise (in the thermodynamic limit), i.e., the critical field can be identified from subleading corrections to the average (over all eigenstates) entanglement entropy.
\end{abstract}

\maketitle

{\it Introduction.} 
Early studies of entanglement entropies in the context of black hole physics~\cite{srednicki_93, holzhey_larsen_94}, quantum information theory~\cite{nielsen_chuang_10}, and the quest for efficient simulation of condensed-matter Hamiltonians~\cite{white_92, schollwoeck_05, schollwoeck_11} rose important questions about the universality of entanglement measures in quantum many-body lattice systems~\cite{amico_fazio_08, peschel_eisler_09, calabrese_cardy_09, eisert_cramer_10}. As a result, several universal features of the bipartite (block) entanglement entropy of {\it ground} states have been identified~\cite{audenaert_eisert_02, osterloh_amico_2002, osborne_nielsen_02, vidal_latorre_03}. Among others, it was established that, in local one-dimensional (1D) fermionic systems (and the spin chains onto which they can be mapped), there is a one-to-one correspondence between criticality (noncriticality) and logarithmic (area law) entanglement entropy scaling~\cite{vidal_latorre_03, latorre_rico_04, hastings_07}. In critical systems described by conformal field theory, the prefactor of the logarithm is the central charge~\cite{vidal_latorre_03, latorre_rico_04, calabrese_cardy_04}.

Subleading terms of the entanglement entropy in many-body ground states can also exhibit universal features. This has been of particular interest in two-dimensional systems. There, a subleading term in the ground state of quadratic fermionic Hamiltonians scaling logarithmically or being a constant distinguishes between critical states with a point-like Fermi surface and noncritical states, respectively~\cite{ding_brayali_08}. (In both cases, the leading term is area law~\cite{wolf_06, gioev_klich_06, barthel_chung_06, li_ding_06, cramer_eisert_07}.) A universal logarithmic subleading term to the leading area law has also been found in some classes of critical states described by conformal field theory~\cite{fradkin_moore_06} and in systems with a spontaneously broken continuous symmetry~\cite{kallin_hastings_11, metlitski_grover_15}. In gapped systems with topological order, a constant correction to the area law may characterize topological properties~\cite{kitaev_preskill_06, levin_wen_06, haque_zozulya_07}.

\begin{figure}[!b]
\centering
\includegraphics[width=0.99\columnwidth]{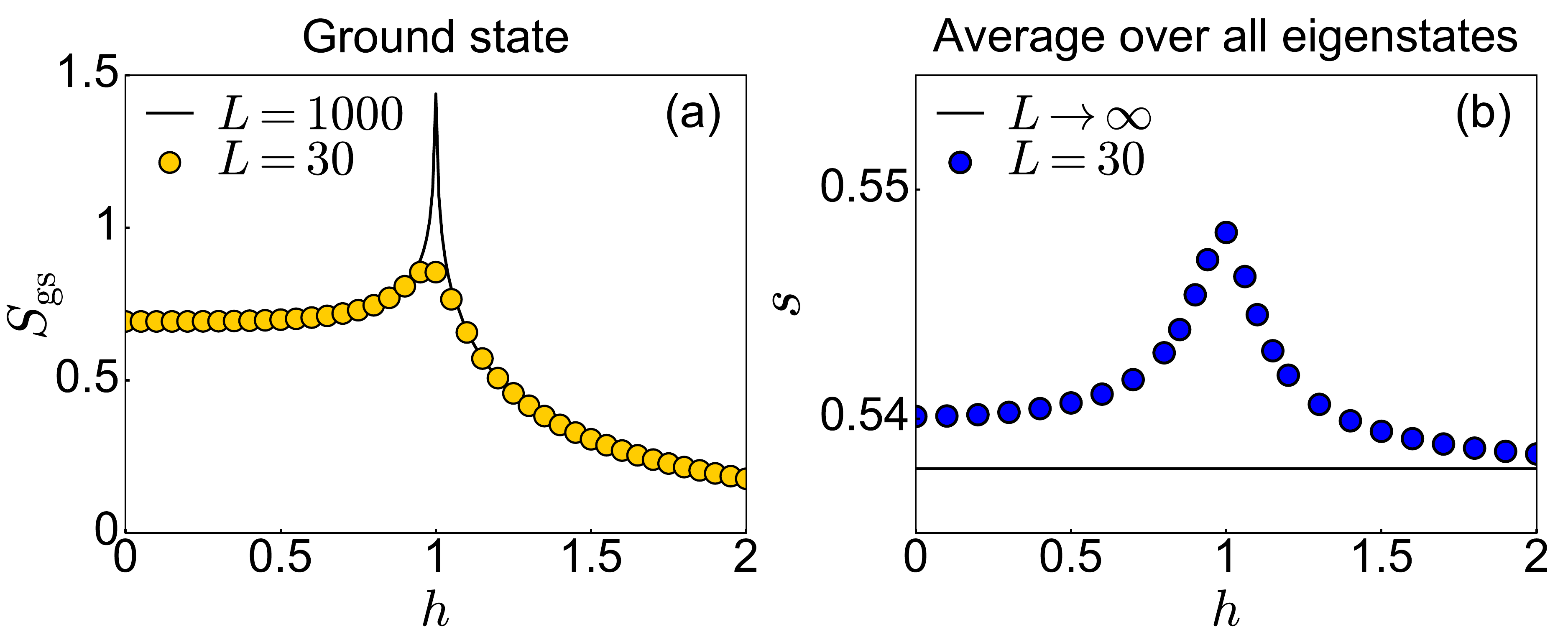}
\caption{Entanglement entropy in the quantum Ising model, Eq.~(\ref{def_Hqi}), as a function of the transverse field $h$. The subsystem volume is one half of that of the system ($f=1/2$). (a) Ground-state entanglement entropy $S_{\rm gs}$. (b) Average (over all eigenstates) entanglement entropy density $s$, defined in Eq.~(\ref{def_savr}). The horizontal line in (b) depicts the extrapolated result in the thermodynamic limit.}
\label{fig1}
\end{figure}

In contrast to ground states, much less is known about the universality (if any) of the entanglement entropy in excited eigenstates of local quadratic Hamiltonians, or of models mappable to them. Recent studies have started exploring the scaling of the entanglement entropy in typical excited eigenstates of a variety of integrable models~\cite{alba09, moelter_barthel_14, storms14, beugeling15, lai15, alba15, nandy_sen_16, vidmar_hackl_17, riddell_mueller_18, zhang_vidmar_18, liu_chen_18}. For quadratic Hamiltonians, it has been shown that typical excited eigenstates exhibit a volume-law scaling. However, they are not maximally entangled if the subsystem volume is a finite fraction of the total volume~\cite{storms14, lai15, nandy_sen_16, vidmar_hackl_17}. The deviation from the maximum is linearly proportional to the volume of the subsystem, and depends on the ratio between the latter and the volume of the system. For translationally invariant models, this was proved by calculating bounds to the average entanglement entropy~\cite{vidmar_hackl_17}.

\begin{figure*}[!t]
\centering
\includegraphics[width=1.95\columnwidth]{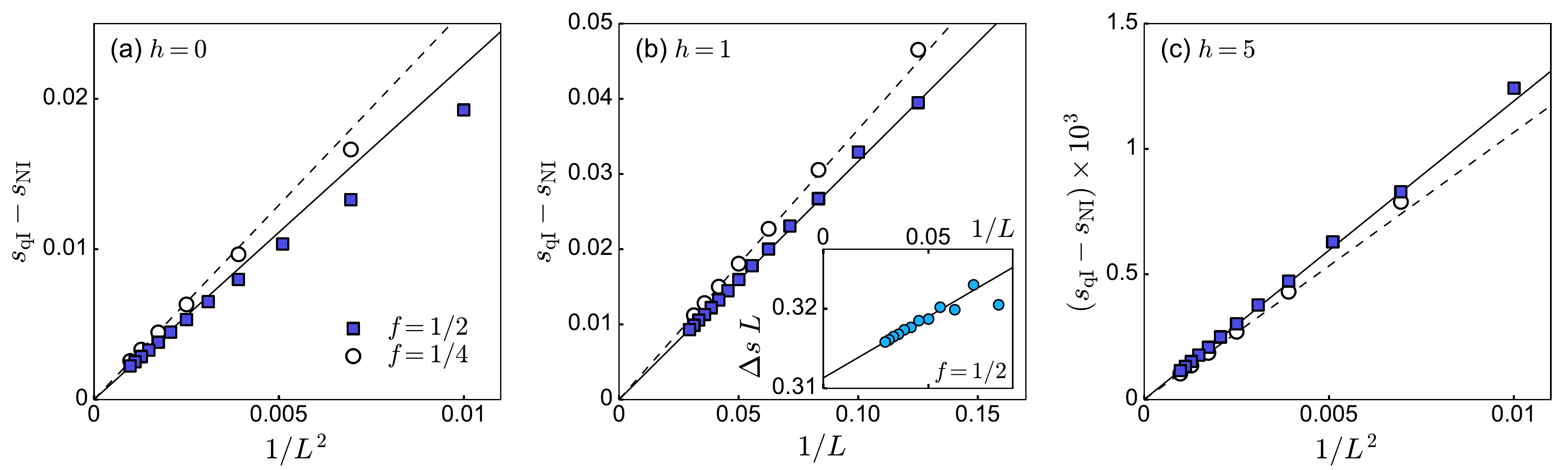}
\caption{Average entanglement entropy density differences, $s_{\rm qI}-s_{\rm NI}$, at two subsystem fractions $f=1/2$ and $1/4$ for: (a) $h=0$, (b) $h=1$, and (c) $h=5$. Lines are fits to the results for $L \geq 24$. Note the difference between the $x$-axes in panels (a) and (c), $1/L^2$, vs panel (b), $1/L$. Inset in (b), rescaled entanglement entropy density $\Delta s \, L = (s_{\rm qI} - s_{\rm NI}) L$ vs $1/L$ at $f=1/2$ and $h=1$. The line depicts a linear fit $A + B/L$ to the results for $L \geq 24$, with $A = 0.311$ and $B = 0.154$.}
\label{fig2}
\end{figure*}

The first goal of this Letter is to study the average eigenstate entanglement entropy of the quantum Ising model in 1D and explore the universality of its leading term. The quantum Ising model has been a paradigmatic model to understand the scaling of ground-state entanglement across a quantum phase transition~\cite{vidal_latorre_03, latorre_rico_04, jin_korepin_04, calabrese_cardy_04, peschel_04, its_jin_05, franchini_its_07, sachdevbook}. As shown in Fig.~\ref{fig1}(a), the ground state entanglement entropy diverges at the critical point (the divergence is logarithmic with the block size), while it is a constant away from criticality. The second goal of this Letter is to determine the subleading term of the average entanglement entropy. Intriguingly, we find (numerically for the average and analytically for its bounds) that the leading correction is a constant at the critical field while it vanishes away from it (in the thermodynamic limit). As a result, the average entanglement entropy density in finite systems [Fig.~\ref{fig1}(b)] looks qualitatively similar to the entanglement entropy of the ground state [Fig.~\ref{fig1}(a)].

{\it Model.}
The quantum Ising Hamiltonian~\cite{pfeuty_70} can be written as
\begin{equation}\label{eq:qI}
\hat H_{\rm qI} = -2J\, \sum_j^L \hat S_j^x \hat S_{j+1}^x - h\, \sum_j^L\hat S_j^z\, ,
\end{equation}
where $\hat S^{x,z}$ are spin-1/2 operators. We use periodic boundary conditions $\hat S^\alpha_{L+1} \equiv \hat S^\alpha_1$. At $h=1$, the ground state exhibits a quantum phase transition between a ferromagnetic phase ($h<1$) and a paramagnetic one ($h>1$). Using the Jordan-Wigner transformation~\cite{jordan_wigner_28, cazalilla_citro_review_11}, one can map the quantum Ising model onto a spinless fermions Hamiltonian (up to a boundary term)
\begin{equation} \label{def_Hqi}
 \hat H_{\rm SF} = - \frac{J}{2} \sum_{j=1}^L \left[ \hat{f}_j^\dagger \hat{f}_{j+1} +  \hat{f}_j^\dagger \hat{f}_{j+1}^\dagger + {\rm H.c.}  \right] - h \sum_{j=1}^L \hat f_j^\dagger \hat f_j \, ,
\end{equation}
where $\hat{f}_j$ ($\hat f_{j+1}^\dagger$) is the fermionic annihilation (creation) operator at site $j$, and $\hat f_{L+1} \equiv \hat f_1$. The Hamiltonian is diagonalized via a Fourier transform $\hat f_j = 1/\sqrt{L}\sum_k e^{ikj} \hat f_k$ and a Bogoliubov transform $\hat f_k = u_k \hat \eta_k - v_k^* \hat \eta_{-k}^\dagger$, which yield $\hat H_{\rm qI} = -(1/2) \sum_k  \varepsilon_k (1-2\hat\eta_k^\dagger \hat\eta_k)$. The single-particle energy is $\varepsilon_k = \sqrt{h^2 + 2hJ\cos{k} + J^2}$ and the coefficients of the Bogoliubov transform are
\begin{equation} \label{def_uk_vk}
u_k = \frac{\varepsilon_k + a_k}{\sqrt{2\varepsilon_k(\varepsilon_k + a_k)}}\, , \;\;\; v_k = \frac{i b_k }{\sqrt{2\varepsilon_k(\varepsilon_k + a_k)}} \, ,
\end{equation}
where $a_k = -J \cos{k} - h$ and $b_k = J \sin{k}$. Many-body eigenstates $|m\rangle$ satisfy $\hat N_k |m\rangle = (1-2\hat \eta_k^\dagger \hat \eta_k)|m\rangle = N_k |m\rangle$, where $N_k = \pm 1$. Note that the Hamiltonian decouples in sectors with even and odd number of particles. The boundary term to Eq.~(\ref{def_Hqi}) results in periodic (antiperiodic) boundary conditions in the odd (even) sector~\cite{suppmat}. Following~\cite{vidmar16}, we treat eigenstates in both sectors exactly. We shall contrast the results for the quantum Ising model to those for noninteracting fermions
\begin{equation} \label{def_Hni}
 \hat H_{\rm NI} = -J \sum_{j=1}^L \left[ \hat{f}_j^\dagger \hat{f}_{j+1} + {\rm H.c.}  \right] \, ,
\end{equation}
onto which the spin-1/2 XX chain can be mapped. We set $J \equiv 1$ in what follows.

{\it Bipartite entanglement entropy.}
We are interested in the von Neumann entanglement entropy of eigenstates of $\hat H_{\rm qI}$, Eq.~(\ref{eq:qI}), for bipartitions of the system into two blocks of length $L_A$ and $L-L_A$. In that case, the entanglement entropy $S_m$ in an eigenstate $|m\rangle$ of $\hat H_{\rm SF}$, Eq.~(\ref{def_Hqi}), is identical to the one in the corresponding eigenstate of $\hat H_{\rm qI}$. $S_m$ can be computed using the fact that the eigenstates $|m\rangle$ of $\hat H_{\rm SF}$ are Gaussian states, i.e., they are fully characterized by the complex structure $[iJ]_A$ (one-body covariance matrix), restricted to the subsystem with $L_A$ sites~\cite{vidal_latorre_03, peschel_03, vidmar_hackl_17}. The latter depends explicitly on $u_k$ and $v_k$. For completeness, in Ref.~\cite{suppmat} we write the explicit expression for $[iJ]_A$ and the corresponding $S_m$.

The average (over all eigenstates) entanglement entropy is $S = 2^{-L} \sum_m S_m$, and we define the average entanglement entropy density as
\begin{equation} \label{def_savr}
 s = \frac{S}{L_A \ln 2} \, .
\end{equation}
Our goal is to determine $s$ of the quantum Ising model ($s_{\rm qI}$) when $L \to \infty$ while $f = L_A/L = {\rm const} > 0$. To this end, we compare $s_{\rm qI}$ to the corresponding average entanglement entropy $s_{\rm NI}$ of the eigenstates of $H_{\rm NI}$, Eq.~(\ref{def_Hni}). For the latter, it was shown that $s_{\rm NI}=0.5378(1)$ at $f = 1/2$ for $L \to \infty$~\cite{vidmar_hackl_17}, with the leading correction vanishing exponentially with $L$. 

In the quantum Ising model, we find large finite-size effects [see Fig.~\ref{fig1}(b)] that are inconsistent with a leading correction decaying exponentially with $L$. Figure~\ref{fig2} shows results for $s_{\rm qI}$, subtracted by $s_{\rm NI}$, for different values of $h$ and for subsystem fractions $f=1/2$ and $1/4$. The results make apparent that the difference scales as $1/L$ at $h=1$, while it scales as $1/L^2$ for $h=0$ and $5$. Based on these results, we make the following conjecture:

\begin{figure}[!t]
\centering
\includegraphics[width=0.99\columnwidth]{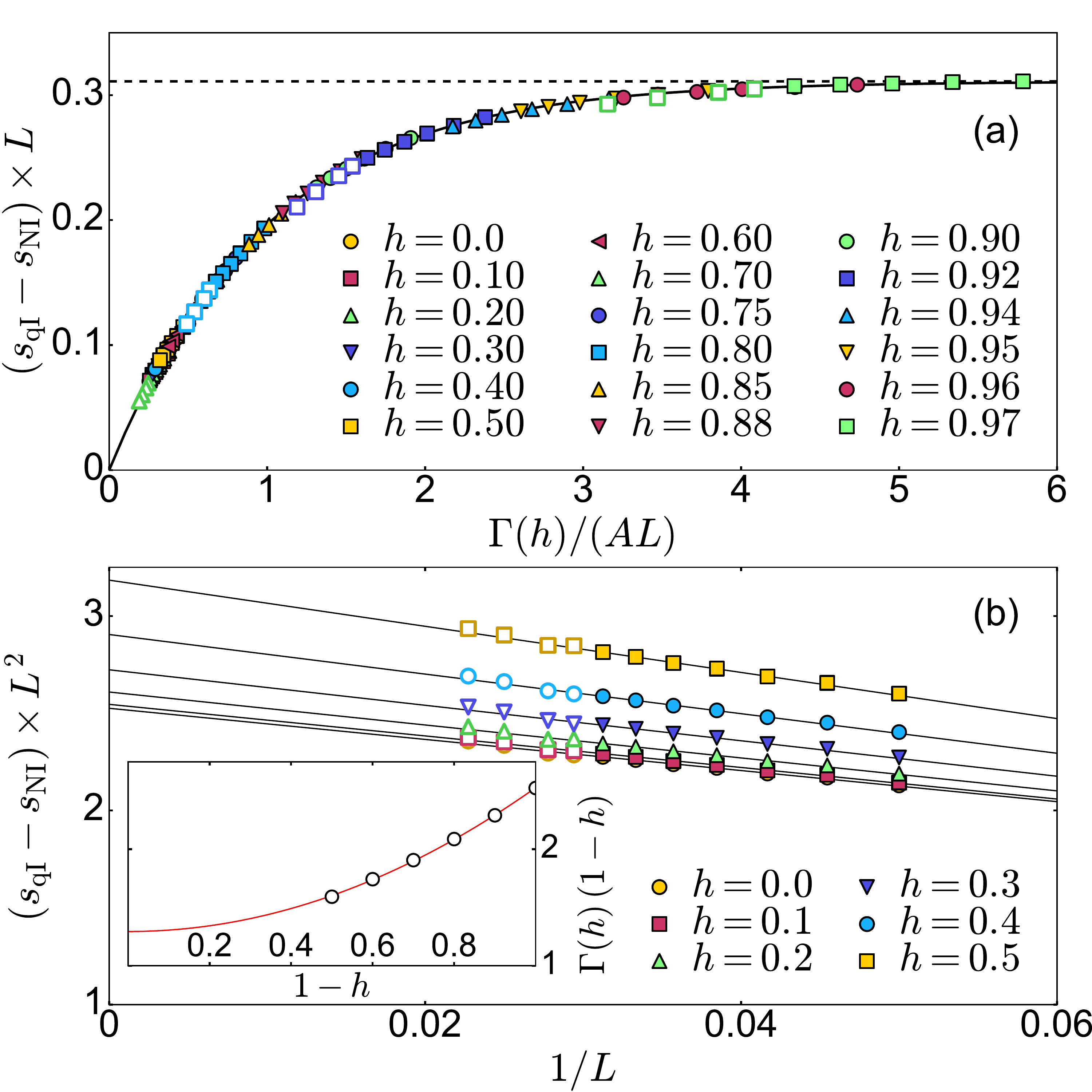}
\caption{(a) Rescaled entanglement entropy density $(s_{\rm qI} - s_{\rm NI}) L$ vs $\Gamma(h)/(AL)$ for $h<1$. Filled symbols are exact results (namely, the average over entire Hilbert space with $2^L$ eigenstates), shown for $22 \leq L \leq 32$, while open symbols show averages over $10^8$ random eigenstates for $34 \leq L \leq 44$. For a given $h$, the color of the open symbols is identical to that of the filled ones. The solid line is the function $A\,[1-e^{-\Gamma(h)/(AL)}]$, where the constant $A = 0.311$ (horizontal line). (b) Rescaled entanglement entropy density $(s_{\rm qI} - s_{\rm NI}) L^2$ vs $1/L$ for $h<1$. The symbol coding is the same as in (a), while the solid lines are linear fits $\Gamma(h) + \zeta(h)/L$ to the exact results for $L \geq 26$. (Inset) The symbols depict the values of $\Gamma(h)$, multiplied by $(1-h)$, plotted vs $(1-h)$. The solid line is a fitting function $\Gamma(h) (1-h) = \alpha (1-h)^2 + \beta$, with $\alpha = 1.234$ and $\beta = 1.294$.}
\label{fig3}
\end{figure}

\begin{figure}[!t]
\centering
\includegraphics[width=0.987\columnwidth]{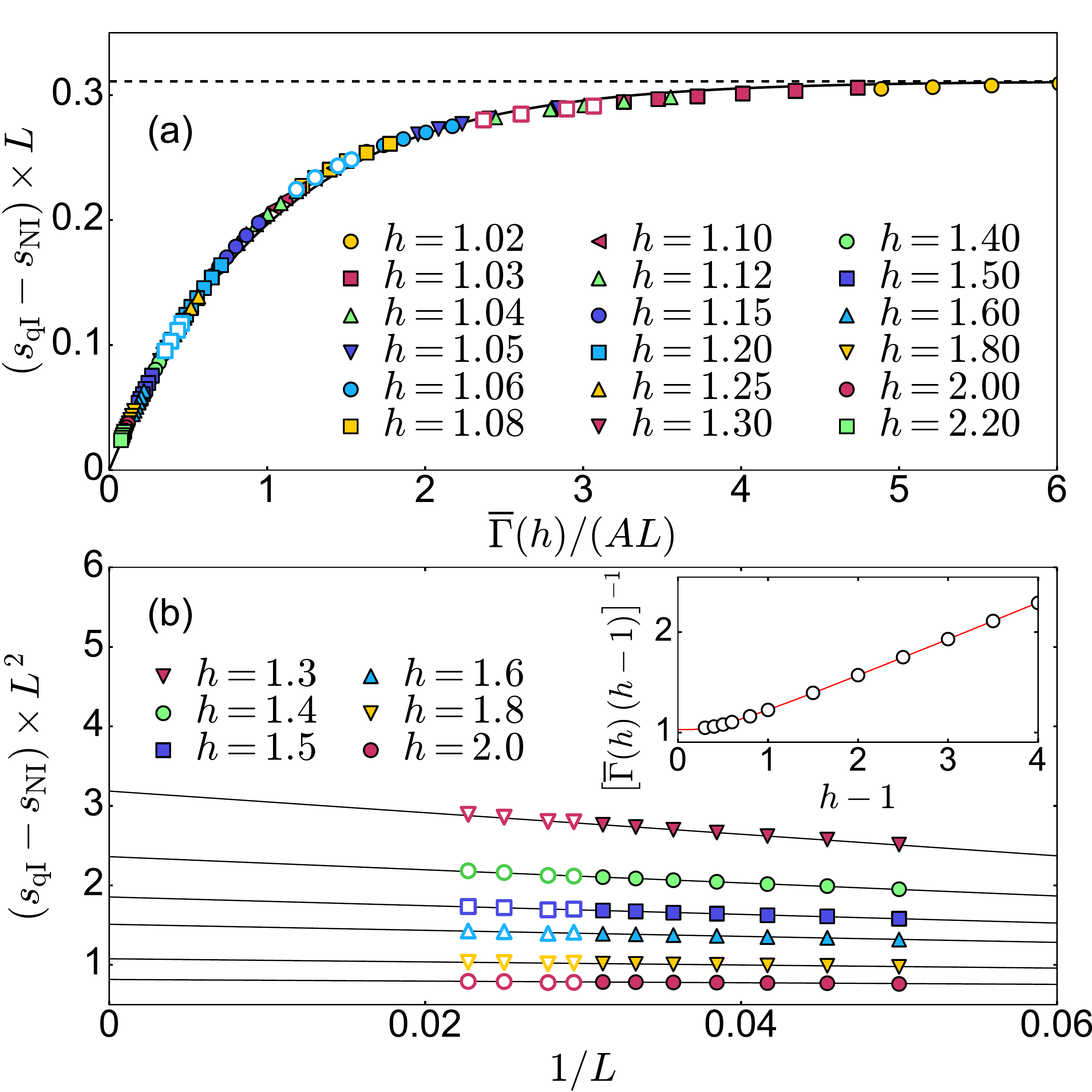}
\caption{(a) Rescaled entanglement entropy density $(s_{\rm qI} - s_{\rm NI}) L$ vs $\overline\Gamma(h)/(AL)$ for $h>1$. The system sizes and color coding are identical to the ones used in Fig.~\ref{fig3}. The solid line is the function $A\,[1-e^{-\overline\Gamma(h)/(AL)}]$, where the constant $A = 0.311$ (horizontal line). (b) Rescaled entanglement entropy density $(s_{\rm qI} - s_{\rm NI}) L^2$ vs $1/L$ for $h>1$. The symbol coding is the same as in (a), while solid lines are linear fits $\overline{\Gamma}(h) + \overline{\zeta}(h)/L$ to the exact results for $L \geq 26$. (Inset) The symbols depict the values of $\overline{\Gamma}(h)$, multiplied by $(h-1)$, plotted vs $(h-1)$. The solid line is a fitting function $\left[\overline{\Gamma}(h) (h-1) \right]^{-1} = \bar\alpha + \bar\beta (h-1) e^{-\bar\eta/(h-1)}$, with $\bar\alpha = 1.028$, $\bar\beta = 0.371$, and $\bar\eta = 0.629$.}
\label{fig4}
\end{figure}

{\bf Conjecture I (Average $s_{\rm qI}$)}.
The leading correction to the average entanglement entropy density difference between the quantum Ising model and noninteracting fermions, for $f>0$, scales as 
\begin{equation} \label{def_conjecture}
 s_{\rm qI} - s_{\rm NI} \propto \left \{
\begin{array}{lcl}
 1/L \,,&& h = 1 \\ 1/L^2\,, && h \neq 1
\end{array}
\right. .
\end{equation}

{\it Numerical test of Conjecture I.}
We test Conjecture I at $f=1/2$ using the exact average entanglement entropies calculated numerically for systems with $L \lesssim 34$. As a first step, we compute the subleading correction at $h=1$ with high precision, $s_{\rm qI}L = s_{\rm NI} L + A$ [see the inset in Fig.~\ref{fig2}(b)] obtaining $A=0.311$. In the next step, we obtain the subleading correction for $h < 1$ ($h>1$) as $s_{\rm qI}L = s_{\rm NI}L + \Gamma(h)/L$ [$s_{\rm qI}L = s_{\rm NI}L + \overline\Gamma(h)/L$] for all values of $h$ for which a high precision finite-size scaling is possible. Examples of such scalings are presented in Figs.~\ref{fig3}(b) and~\ref{fig4}(b) for $h<1$ and $h>1$, respectively. We find that the functions $\Gamma(h)$ and $\overline\Gamma(h)$ diverge at $h=1$. The insets in Figs.~\ref{fig3}(b) and~\ref{fig4}(b) show fits to $\Gamma(h)$ and $\overline\Gamma(h)$, respectively, in the entire regime of $h$. The most important property for the analysis that follows is that $\Gamma(h) (1-h)$ and $\overline\Gamma(h) (h-1)$ are functions that are both smooth about $h = 1$. As a result, $\lim_{h\to 1^-} \Gamma(h) \to \infty$ and $\lim_{h \to 1^+} \overline{\Gamma}(h) \to \infty$.

Our main results, following from the previous calculations, are shown in Figs.~\ref{fig3}(a) and~\ref{fig4}(a). They reveal that $(s_{\rm qI} - s_{\rm NI})L$ is, for finite $L$, a universal function of $\Gamma(h)/L$ for $h<1$ and of $\overline{\Gamma}(h)/L$ for $h>1$. This uncovers the scaling when $L \to \infty$. Whenever $h \neq 1$, $\lim_{L\to\infty} \Gamma(h)/L \to 0$ and $\lim_{L\to\infty} \overline\Gamma(h)/L \to 0$, so that $s_{\rm qI} \to s_{\rm NI}$ with a correction that is, at most, $O(L^{-2})$. 

Moreover, the results in Figs.~\ref{fig3}(a) and~\ref{fig4}(a) allow us to identify that the scaling function is close to
\begin{equation} \label{fss_scaling}
 (s_{\rm qI} - s_{\rm NI})L = A\,\left[1-e^{-\frac{\Gamma(h)}{A  L}}\right]
\end{equation}
for $h<1$, and similarly for $h>1$ upon replacing $\Gamma(h) \to \overline{\Gamma}(h)$. The functions are shown as solid lines in Figs.~\ref{fig3}(a) and~\ref{fig4}(a). They describe both the critical and noncritical regime of $h$ when $L\to\infty$, namely, $(s_{\rm qI} - s_{\rm NI})L \to A$ if $h=1$ and $(s_{\rm qI} - s_{\rm NI})L \to \Gamma(h) /L$ if $h \neq 1$.

{\bf Corollary I}.
The average entanglement entropy density of the quantum Ising model in the thermodynamic limit is $s_{\rm qI} = s_{\rm NI}$ for all values of the transverse field $h$.

{\bf Corollary II}.
The average entanglement entropy of the quantum Ising model for $L\gg1$ can be written as
\begin{equation}
 S_{\rm qI} = S_{\rm NI} + \delta_{h,1} \, {\rm const.} + O(1/L) \, ,
\end{equation}
i.e., the subleading term is a constant in the thermodynamic limit if and only if $h=1$.

{\it Analytical results for the bounds.}
Next, we compute the exact bounds for the average. We use that: $L_A \ln 2 - \frac{\langle {\rm Tr}[iJ]_A^2 \rangle}{2} \ln 2 \leq S_{\rm qI} \leq L_A \ln 2 - \frac{\langle {\rm Tr}[iJ]_A^2 \rangle}{4}$~\cite{vidmar_hackl_17}, where
\begin{equation} \label{def_tr2}
\langle {\rm Tr}[iJ]_A^2 \rangle = 2L_A f - \frac{2}{L^2} \sum_k 4 |u_k|^2 |v_k|^2 \frac{\sin^2(L_A k)}{\sin^2(k)} \,
\end{equation}
is the spectral average of the trace of the square of $[iJ]_A$. Note that the allowed values of $k$ in the sum are determined by the boundary conditions in each sector. To evaluate Eq.~(\ref{def_tr2}), we use that $\langle {\rm Tr}[iJ]_A^2 \rangle = (\langle {\rm Tr}[iJ]_A^2 \rangle_{\rm p} + \langle {\rm Tr}[iJ]_A^2 \rangle_{\rm a})/2$, where $\langle\cdot\rangle_{\rm p}$ ($\langle\cdot\rangle_{\rm a}$) is the spectral average over all eigenstates for periodic ``p'' [antiperiodic ``a''] boundary conditions.

\begin{figure}[!t]
\centering
\includegraphics[width=0.99\columnwidth]{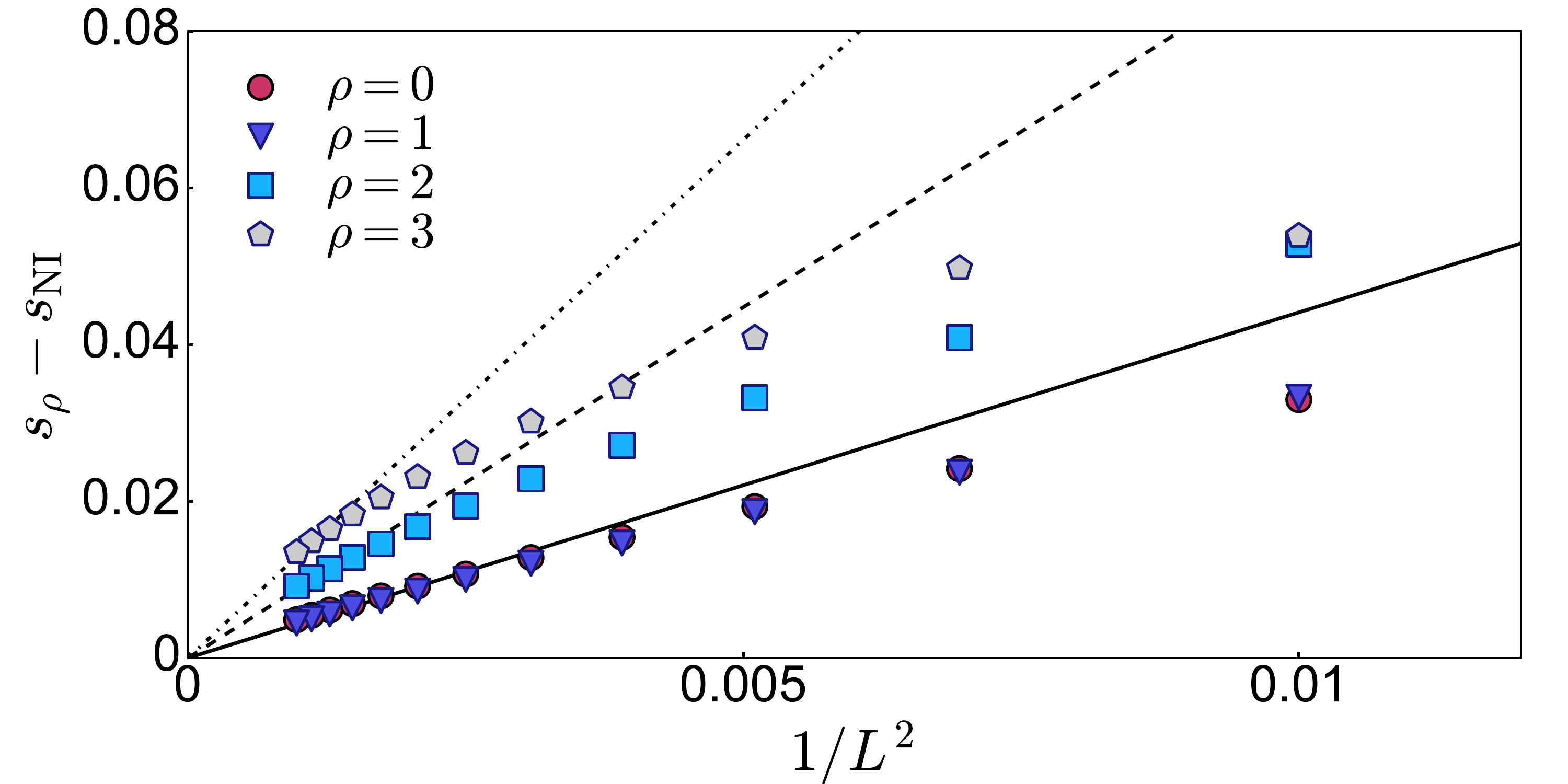}
\caption{Rescaled entanglement entropy density $s_\rho - s_{\rm NI}$ vs $1/L^2$ for various translationally invariant quadratic models. We replace $u_k$ and $v_k$ in Eq.~(\ref{def_uk_vk}) by $u_k = \cos[\sin(\rho k)]$ and $v_k = \sin[\sin(\rho k)]$ for $\rho \geq 1$, and $u_k = \cos(k)$ and $v_k = \sin(k)$ for $\rho = 0$. Solid and dashed lines are linear fits $w_\rho/L^2$ to the results for $L \geq 26$, with $w_0 = w_1 = 4.41$ and $w_2 = 8.96$. The dashed-dotted line is a guide to the eye with $w_3 = 3 w_1$.}
\label{fig5}
\end{figure}

Using $|u_k|^2$ and $|v_k|^2$ from Eq.~(\ref{def_uk_vk}) one gets that, for $h=1$, the addends in the second term of Eq.~(\ref{def_tr2}) can be written as $(1/2)\sin^2(L_A k)/[1+\cos(k)]$, which yields~\cite{suppmat}
\begin{equation} \label{trace_h1}
 \langle {\rm Tr}[iJ]_A^2 \rangle = 2L_A f - f (1-f) \, ,
\end{equation}
i.e., the correction to the volume-law term is a constant. On the other hand, for $h \neq 1$, one can set an upper bound to the second term in Eq.~(\ref{def_tr2}) by replacing $\sin^2(L_A k)$ with 1 and $|u_k|^2 |v_k|^2 /\sin^2(k)$ with $1/(h-1)^2$. This yields
\begin{equation}
 \langle {\rm Tr}[iJ]_A^2 \rangle > 2L_A f - \frac{1}{L} \frac{8}{(h-1)^2} \, .
\end{equation}
These results for the bounds are consistent with our numerical results for the average. They also identify a quantity related to $S_{\rm qI}$, namely, ${\rm Tr}[iJ]_A^2$, for which the spectral average exhibits a constant correction to the leading volume-law term only at the critical field.

{\it Universality.} Having found that the leading term in $S_{\rm qI}$ and its first order bounds are identical to the ones for noninteracting fermions, we conjecture:

{\bf Conjecture II (Universality of $S$)}.
The leading (volume-law) term in the average entanglement entropy is identical for all translationally invariant quadratic fermionic Hamiltonians.

{\it Numerical test of Conjecture II.}
We replace the coefficients $u_k$ and $v_k$ of the Bogoliubov transform, Eq.~(\ref{def_uk_vk}), by functions that are consistent with translational invariance ($|v_k|^2+|u_k|^2=1$, $u_k=u_{-k}$, and $v_k=-v_{-k}$). Results for the average entanglement entropy density $s_\rho$ for four such choices of $u_k$ and $v_k$ are shown in Fig.~\ref{fig5}. Remarkably, in all cases we observe that $s_\rho$ approaches $s_{\rm NI}$. The subleading correction is $\propto L^{-2}$, i.e., identical to the one in the quantum Ising model away from criticality.

{\it Summary and discussion.}
We argued that the leading term in the average entanglement entropy of the quantum Ising model is identical to that of noninteracting fermions, which, in turn, we conjecture is universal for translationally invariant quadratic fermionic Hamiltonians. Such models appear to belong to a different ``universality class'' when compared to quadratic fermionic Hamiltonians described by random matrices, for which the leading term in the average entanglement entropy is different~\cite{liu_chen_18}, and to models with extended unit cells, for which previous numerical work~\cite{storms14} revealed larger average entanglement entropies than those reported here.

We also studied the corrections to the leading term. We showed that in the quantum Ising model they allow one to identify the critical field for the quantum phase transition. The correction is order one at the critical field and vanishes otherwise in the thermodynamic limit. The fact that the correction depends on whether the field is at the critical value for the quantum phase transition or away from it is unexpected considering that the average entanglement entropy is dominated by states in the middle of the spectrum (at ``infinite temperature''). It highlights the need for studies of the average entanglement entropy in other models with quantum phase transitions. Our results may be of relevance to periodically kicked Ising systems, for which a divergence of correlation functions was observed in averages over all eigenstates of a Floquet Hamiltonian at the critical field~\cite{prosen_00}.

\medskip

{\it Acknowledgments.}
We thank X. Chen and D. Iyer for discussions. We acknowledge support from the Slovenian Research Agency, research core funding No.~P1-0044 (L.V.), a Mebus Fellowship (L.H.), the Max Planck Harvard Research Center for Quantum Optics (L.H.), and the National Science Foundation Grant Nos.~PHY-1748958 (L.V. and M.R.),~PHY-1806428 (E.B.) and~PHY-1707482 (M.R.). The computations were done at the Institute for CyberScience at Penn State.

\bibliographystyle{biblev1}
\bibliography{references}

\newpage
\phantom{a}
\newpage
\setcounter{figure}{0}
\setcounter{equation}{0}

\renewcommand{\thetable}{S\arabic{table}}
\renewcommand{\thefigure}{S\arabic{figure}}
\renewcommand{\theequation}{S\arabic{equation}}

\renewcommand{\thesection}{S\arabic{section}}

\onecolumngrid

\begin{center}

{\large \bf Supplemental Material:\\
Volume Law and Quantum Criticality in the \\ Entanglement Entropy of Excited Eigenstates of the Quantum Ising Model
}\\

\vspace{0.3cm}

Lev Vidmar,$^{1,2}$ Lucas Hackl,$^{3,4,5}$ Eugenio Bianchi$^{4,5}$, Marcos Rigol$^{4,2}$\\
$^1${\it Department of Theoretical Physics, J. Stefan Institute, SI-1000 Ljubljana, Slovenia}\\
$^2${\it Kavli Institute for Theoretical Physics, University of California, Santa Barbara, California 93106, USA}\\
$^3${\it Max  Planck  Institute of Quantum Optics, Hans-Kopfermann-Stra\ss e 1, D-85748 Garching bei M\"unchen, Germany}\\
$^4${\it Department of Physics, The Pennsylvania State University, University Park, PA 16802, USA}\\
$^5${\it {Institute for Gravitation and the Cosmos, The Pennsylvania State University, University Park, PA 16802, USA}}

\end{center}

\vspace{0.6cm}

\twocolumngrid

\label{pagesupp}

\section{Entanglement entropy of an eigenstate} \label{app1}

All the properties of eigenstates of quadratic models are encoded in $L\times L$ one-body correlation matrices. They form a $2L\times2L$ matrix $iJ$, which is a linear complex structure
\begin{align}\label{eq:lcs}
iJ&=\left(\begin{array}{c|c}
\langle m|\hat f_i^\dagger \hat f^{}_j - \hat f^{}_j \hat f_i^\dagger|m\rangle & \langle m|\hat f_i^\dagger \hat f_j^\dagger - \hat f_j^\dagger \hat f_i^\dagger|m\rangle\\ \hline \langle m|\hat f^{}_i \hat f^{}_j - \hat f^{}_j \hat f^{}_i|m\rangle & \langle m| \hat f^{}_i \hat f_j^\dagger - \hat f_j^\dagger \hat f^{}_i|m\rangle
\end{array}\right) .
\end{align}

In the quantum Ising model, eigenstates belong either to the even or the odd particle-number sector. Each sector has a set of allowed $k$ vectors, which we denote as ${\cal K}^{+} $ (even sector) and ${\cal K}^{-}$ (odd sector) \cite{vidmar16}. 

The matrix elements of $iJ$ in Eq.~(\ref{eq:lcs}) are:\\
(i) If $|m\rangle$ belongs to the even sector
\begin{align}
 \langle m | \hat f_j^\dagger \hat f^{}_l |m\rangle = & - \frac{1}{L} \sum_{k \in {\cal K}^{+}} N_k \cos[k(j-l)] u_k^2 \nonumber \\
 & + \frac{1}{2L} \sum_{k \in {\cal K}^{+}} N_k e^{ik(j-l)} + \frac{1}{2} \delta_{j,l}
\end{align}
and
\begin{equation}
 \langle m | \hat f_j^\dagger \hat f_l^\dagger |m\rangle = \frac{i}{L} \sum_{k \in {\cal K}^{+}} N_k \sin[k(j-l)] u_k v_k \, ,
\end{equation}
where ${\cal K}^{+} = \{ \pi/L + n 2\pi/L \; | \; n = 0, ..., L/2-1 \}$.

\noindent (ii) If $|m\rangle$ belongs to the odd sector
\begin{align} \label{def_matele_odd1}
 \langle m | \hat f_j^\dagger \hat f^{}_l |m\rangle = & - \frac{1}{L} \sum_{k \in {\cal K}^{-}} N_k \cos[k(j-l)] u_k^2 \nonumber \\
 & + \frac{1}{2L} \sum_{k \in {\cal K}^{-} \backslash \{ 0,\pi \}} N_k e^{ik(j-l)} + \frac{1}{2} \delta_{j,l}
\end{align}
and
\begin{equation} \label{def_matele_odd2}
 \langle m | \hat f_j^\dagger \hat f_l^\dagger |m\rangle = \frac{i}{L} \sum_{k \in {\cal K}^{-} \backslash \{ 0,\pi \}} N_k \sin[k(j-l)] u_k v_k \, ,
\end{equation}
where ${\cal K}^{-} = \{ n 2\pi/L \; | \; n = 0, ..., L/2-1 \}$. Note that in two of the three sums over $k$ in Eqs.~(\ref{def_matele_odd1}) and~(\ref{def_matele_odd1}), the vectors $0$ and $\pi$ are excluded from the sum.

Correlations of a subsystem $A$ containing $L_A$ sites are encoded in the restricted complex structure $[iJ ]_{A}$, the $2L_A \times 2L_A$ matrix obtained by restricting the matrix $iJ$ in Eq.~(\ref{eq:lcs}) to the entries with $j,l\in A$. The entanglement entropy of subsystem $A$ in eigenstate $|m\rangle$ can be computed as~\cite{vidmar_hackl_17}
\begin{align}\label{def_Salpha}
S_m = - \mathrm{Tr} \left\{ \left(\frac{1\!\!1+[iJ]_A}{2}\right)\ln \left(\frac{1\!\!1+[iJ]_A}{2}\right) \right\}.
\end{align}
We diagonalize the matrix $[iJ]_A$ numerically for each eigenstate to calculate $S_m$, and then average over all eigenstates $|m\rangle$ to obtain the spectral average $S$ that is reported in the main text.

\section{Derivation of Eq.~(\ref{trace_h1})} \label{app2}

Since we express the spectral average of ${\rm Tr}[iJ]_A^2$ in the quantum Ising model as the mean of spectral averages over all eigenstates with periodic (using $k \in {\cal K}^{-}$) and antiperiodic (using $k \in {\cal K}^{+}$) boundary conditions, we can express Eq.~(\ref{def_tr2}) at $h=1$ as
\begin{equation} \label{def_trace2}
\langle {\rm Tr}[iJ]_A^2 \rangle = 2L_A f - \frac{1}{L^2} \sum_{k \in {\cal K}^{+} \cup {\cal K}^{-} \backslash \{ \pi \}} \frac{1}{2} \frac{\sin^2(L_A k)}{[1+\cos(k)]} \, .
\end{equation}
Here, $k=\pi$ is excluded from the sum since $u_{\pi} = 0$~\cite{vidmar16}.
By inserting $k=\pi$ back to the sum in Eq.~(\ref{def_trace2}) we get
\begin{equation} \label{def_trace3}
\langle {\rm Tr}[iJ]_A^2 \rangle = 2L_A f - \frac{1}{L^2} \sum_{k \in {\cal K}^{+} \cup {\cal K}^{-} } \frac{1}{2} \frac{\sin^2(L_A k)}{[1+\cos(k)]} + f^2 \, .
\end{equation}
Moreover, realizing that
\begin{equation}
 \frac{1}{2L}  \sum_{k \in {\cal K}^{+} \cup {\cal K}^{-} } \frac{\sin^2(L_A k)}{[1+\cos(k)]}  = L_A \, ,
\end{equation}
we arrive at Eq.~(\ref{trace_h1}) in the main text.

\end{document}